\documentclass[twocolumn,amsmath,amssymb,pre]{revtex4}
\usepackage{graphicx,color,longtable}

\begin{document}

\makeatletter

\title{On measuring colloidal volume fractions}
\author{Wilson C. K.  Poon}
\affiliation{Scottish Universities Physics Alliance (SUPA) and The School
of Physics, University
of Edinburgh, Kings Buildings, Mayfield Road, Edinburgh EH9 3JZ,
U. K.}

\author{Eric R. Weeks}
\affiliation{Department of Physics, Emory University, Atlanta, GA
30322 U.S.A.}

\author{C. Patrick Royall}
\affiliation{School of Chemistry, University of Bristol, Bristol, BS8
1TS, U.K}

\begin{abstract}
Hard-sphere colloids are popular as models for testing fundamental theories in condensed matter and statistical physics, from crystal nucleation to the glass transition. A single parameter, the volume fraction ($\phi$), characterizes an ideal, monodisperse hard-sphere suspension. In comparing experiments with theories and simulation, researchers to date have paid little attention to likely uncertainties in experimentally-quoted $\phi$ values. We critically review the experimental measurement of $\phi$ in hard-sphere colloids, and show that while statistical uncertainties in comparing relative values of $\phi$ can be as low as $10^{-4}$, systematic errors of 3-6\% are probably unavoidable. The consequences of this are illustrated by way of a case study comparing literature data sets on hard-sphere viscosity and diffusion.
\end{abstract}
\vspace{0.5cm}
  
\maketitle



\section{Introduction}

Excluded volume effects dominate the behaviour of liquids around the triple point \cite{Widom67} and play a key role in structuring crystalline \cite{EvansBook64} and amorphous solids \cite{Zallen98}. Thus, hard spheres have long functioned as a reference system for theoretical and simulational studies of condensed matter. In 1986, Pusey and van Megen demonstrated \cite{pusey86} that suspensions of sterically-stabilised polymethylmethacrylate (PMMA) colloids showed nearly-perfect hard-sphere equilibrium phase behaviour, undergoing a first-order phase transition from a fluid to a crystalline state at concentrations around those predicted some time ago by computer simulations for hard spherical particles \cite{Hoover67,Hoover68}. Soon afterward, the same authors showed \cite{Pusey87a} that PMMA colloids underwent a glass transition at even higher concentrations. Subsequently, in a 1991 review, Pusey enunciated the `colloids as atoms' paradigm \cite{Pusey91r} --- Brownian suspensions can be used as `test tube simulations' of many generic condensed matter phenomena such as crystallization and vitrification. Since then, the use of colloids as models has become very popular, especially since the addition of non-adsorbing polymers can be used to induce an inter-particle attraction `tuneable' separately in its range and depth \cite{Poon02r,HenkBook}. In such colloid-polymer mixtures, gas-liquid coexistence can be studied \cite{lekkerkerker92,ilett95}, including interfaces and criticality \cite{aarts04,royall2007nphys}, as well as novel modes of arrest \cite{PhamScience2002}. Due to their slow intrinsic time scales, colloids are also ideal models for studying phase transition kinetics \cite{Lekkerkerker02r}.

While a range of inter-particle interactions are now available in model colloids, hard spheres remain an important reference system for which very direct comparison between experiments and theoretical calculations or computer simulations is in principle possible. The behaviour of a single-sized, or monodisperse, system of hard spheres is controlled by one parameter, the volume fraction $\phi$, i.e. the fraction of the total volume $V$ that is filled by $N$ spheres, each of radius $a$, 
\begin{equation}
\phi = \frac{4}{3}\pi a^3 \frac{N}{V}. \label{phidef}
\end{equation}
Since $\phi$ is precisely known in theory or simulations, a comparison with experiments is straightforward provided that this quantity is also accurately measurable for real suspensions. Much of the literature has indeed proceeded on this basis, assuming that $\phi$ is unproblematically known from experiments. 

However, as Pusey and van Megen pointed out in a symposium article \cite{Pusey87b} following their {\it Nature} paper,\cite{pusey86} the experimental determination of $\phi$ is emphatically {\em not} unproblematic, because: (1) no real colloid is truly `hard', since there is always some softness in the interparticle potential; and (2) real colloids always have a finite size distribution, i.e. they are polydisperse. Thus, Pusey and van Megen calculated an experimental `effective' hard-sphere volume fraction $\phi_{\rm E}$, and found that the freezing and melting volume fractions of their system were $0.494$ and $0.535$ respectively, compared to 0.494 and 0.545 in simulations.\cite{Hoover67,Hoover68} Their careful conclusion reads: `Despite ambiguities \ldots in the experimental determination of the coexistence region this difference is probably significant.' 

The same degree of caution has not characterized the literature since. Experimental reports typically do not discuss in any detail the method used for arriving at $\phi$. On the other hand, theory or simulations almost always take experimental reports of $\phi$ at face value and proceed to use the data on this basis. This is unsatisfactory, particularly in situations where theory testing demands a degree of accuracy and certainty in the experimental $\phi$ that is probably unattainable. In this paper, we critically review a plethora of methods for the experimental determination of $\phi$ in hard-sphere suspensions, evaluate the degree of accuracy attainable in each case, comment on the potential discrepancies between methods, and give a case study showing how different experiments should be compared taking into account possible difference in $\phi$ determination. 

With the increasing popularity of confocal microscopy, direct counting of particles is becoming a standard method for determining $\phi$ (see Sec.~\ref{particleCounting}). This method depends on knowing the particle size. Thus, after introducing model colloids (Section~\ref{particles}), we review particle sizing (Section~\ref{sizesec}). The `classic' method for determining $\phi$ is via the crystallization phase behaviour, which changes with polydispersity\cite{fasolo03,Sollich10a}. So we review polydispersity measurements (Section~\ref{polysec}) before turning to consider the determination of $\phi$ in detail (Section~\ref{volsec}). We finish with a case study (Section~\ref{casestudy}) and a Conclusion.

\section{The particles}
\label{particles}

We focus on suspensions of nearly-perfect hard spheres. 
A system that potentially behaves most like perfect hard spheres is charge-stabilised colloidal silica in water. When the charges are sufficiently screened out by the addition of salt \cite{piazza1993,shikata94,cheng02}, the resulting suspension is very close to being hard-sphere-like. However, a significant drawback of silica as a model system is that these particles have a density $\rho \approx 2.2$ g/cm$^3$, and it has proved impossible to find solvents that match this density. The sedimentation problem can be alleviated by using smaller particles, but at the expense of increasing polydispersity. Charged-stabilised polystyrene spheres are also potentially very hard-sphere-like, and are easier to density match, though (unlike silica) almost impossible to refractive index match. 

A more popular model hard sphere system is PMMA, sterically stabilized by a $\delta \lesssim 10$~nm layer of PHSA (poly-12-hydroxystearic acid) \cite{antl86}. This layer confers a degree of softness to the interparticle potential on the scale of $\delta$. PMMA particles can be dispersed in solvent mixtures that match both the particles' density and index of refraction \cite{segre01,dinsmore01}.  Unfortunately, particle swelling by solvents is endemic \cite{bosma02,ohtsuka08}. Furthermore, the swelling process can take several weeks, so the particle size changes over time, though heat shock may speed up the process to taking only a few hours \cite{kaufman06}.  As swelling is poorly characterized, {\em in situ} measurement is the only reliable means of ensuring that it is complete before the particles are used in experiments. This is particularly important at high $\phi$, where many properties are steep functions of the concentration: an $x$\% increase in the particle radius translates into $\gtrsim 3x$\% in $\phi$. Thus, e.g., an index-matching mixture of cis-decalin and tetrachloroethylene causes 20\% swelling, which has a drastic effect upon the colloid volume fraction \cite{ohtsuka08}. Finally, batch-to-batch variations generate further uncertainties. 

Below, we focus on PMMA particles, although much of what we say will also apply to silica and other model systems. 

\section{Measuring size}
\label{sizesec}

The basic parameter characterizing spherical colloidal particles is their radius, $a$. Some methods of determining $\phi$, most obviously by `counting' from confocal microscopy images, depend directly on measuring $a$. In this section, we critically review the measurement of particle size.

\begin{figure}
\begin{center}
\includegraphics[width=6cm]{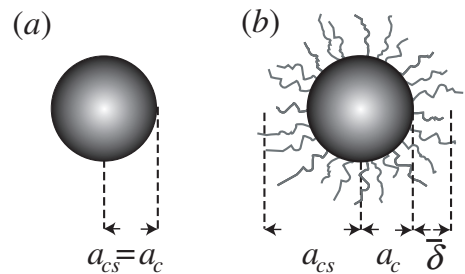}
\end{center}
\caption{
Definitions of particle radii. (a) Non-sterically-stabilized particles, e.g. silica, with a core radius $a_c$. (b) Sterically-stabilized particle, with surface `hairs' (not drawn to scale), where additionally the average hair thickness $\bar{\delta}$ and the core-shall radius $a_{CS} = a_C + \bar{\delta}$ are needed for a full characterisation.} 
\label{pmma}
\end{figure}

\subsection{Definition of particle radius}
\label{radiusdefine}

At first sight, defining `the particle radius' should be simple in a nearly-monodisperse colloid. But this is deceptive, Fig.~\ref{pmma}. For sterically-stabilized  particles like PMMA, we can usefully define as least four different radii. First, the {\em hydrodynamic} radius, $a_H$, occurs in the relationship between the drag, $f$, on a particle moving at velocity $v$ in a fluid of viscosity $\eta$ at low Reynolds numbers, $f = \xi v$, where the friction coefficient 
\begin{equation}
\xi = 6\pi \eta a_H. \label{xi}
\end{equation} 
Relating $a_H$ to $\phi$ is a non-trivial problem in hydrodynamics. Somewhat more directly related to $\phi$ is the {\em core} radius, $a_C$, which is the radius of the sterically-stabilized particles {\em minus} the stabilizing hairs. Thirdly, if we can determine the average hair thickness $\bar{\delta}$, then the {\em core-shell} radius, $a_{CS} = a_C + \bar{\delta}$. For `hairless' particles, such as charge-stabilized silica, $a_{CS} = a_C$. Finally, we may assign an {\em effective hard sphere} radius, $a_{\rm eff}$, to the particles to obtain the best fit to theory or simulations of hard-sphere behaviour in a certain range of $\phi$, so that $a_{\rm eff}$ is inevitably dependent on the chosen property and $\phi$ range. 

We will not review established sizing methods in any detail, but will reference existing literature and note cautionary points. Then we will introduce a number of newer methods. 

\subsection{Measuring radius: established methods}
\label{established}

Scattering methods have a long history in sizing spherical particles\cite{Bombannes,berne76,Chu07}. Static and dynamic scattering determine the size of particles by measuring the time-averaged or fluctuating intensity of the scattered light respectively. 

Dynamic light scattering (DLS) and its X ray equivalent, X ray photon correlation spectroscopy (XPCS), measure the diffusion coefficient of particles, which is related to the friction coefficient via  the Stokes-Einstein-Sutherland relation\cite{sutherland1905,einstein1905a,einstein1906a}: $D = k_B T/\xi$. DLS and XPCS therefore determine $a_H$ (cf. Eqn.~\ref{xi}), and are most useful in the case of particles consisting of core only, such as silica, since it is less clear how to relate $a_H$ to $a_{CS}$ for core-shell particles such as PMMA. Note that the accuracy of this method depends on having an accurate value for $\eta$, the solvent viscosity, which is temperature dependent. For example, we have found that for the common solvent mixture cyclohexylbromide and decalin
(85\%/15\% by weight) $24^\circ$C, $\eta =
2.120$~mPa$\cdot$s and $d \eta / dT=-0.029$ mPa$\cdot$s/K.  Thus,
a 1$^\circ$C uncertainly in $T$ is a 0.3\% uncertainty of $T$ but
a 1.7\% uncertainty in $T/\eta$ and therefore in $a_H$.

Static light scattering (SLS), small-angle X ray scattering (SAXS) or small-angle neutron scattering (SANS) can potentially determine $a_C$ and $a_{CS}$. Since the core and shell of (say) a PMMA particle in general has different contrasts to light, X rays (refractive index, $n$, in both cases) and neutrons (scattering length, $b$), the diffraction pattern of a single particle is determined by the interference of radiation scattered from these two parts. Fitting this diffraction pattern (the form factor) therefore can in principle yield $a_C$ and $a_{CS} = a_C + \delta$. In a solvent with $n$ or $b$ quite different from both the core and the shell, the whole entity scatters more or less as a homogeneous sphere and a radius close to $a_{CS}$ is returned from form-factor fitting. When solvent mixtures are  used to `tune' the relative contrasts of core and shell, even a small amount of a minority component in the solvent mixture can swell the particles by up to 10\% or more\cite{bosma02,ohtsuka08}, and the fractional swelling of core and shell is not necessarily identical. In XPCS, where the shell has little contrast, $\delta$ cannot be accurately determined; however, the brightness of the beam gives many orders of oscillations in the form factor, allowing very accurate data fitting. 

For both static and dynamic scattering, samples must be dilute enough so that the properties of non-interacting particles are measured in the single-scattering limit. The only sure way to know that this has been achieved is to collect data at different $\phi$ and look for the convergence in the $\phi \rightarrow 0$ limit. Static scattering at finite $\phi$ gives the static structure factor as a function of scattering vector, $S(q)$. Fitting this to, e.g., the Percus-Yevick form \cite{Hansen} or simulations yields simultaneously $a_{\rm eff}$ and $\phi$, although polydispersity is a significant complication\cite{Vrij1988}. Alternatively, the Bragg peaks in $S(q)$ from colloidal crystals at fluid-crystal coexistence  can be used to deduce $a_{\rm eff}$ if the melting point is known (but see Section~\ref{polysec} for caveats). 

Electron microscopy (EM) measures $a_C$ of dried particles, because drying collapses the steric-stabilizing `hairs' in core-shell particles such as PMMA, and deswells particles swollen by solvent when dispersed. Various optical microscopies can, in principle, be used in the same way as EM for sizing particles; caveats are pointed out in Section~\ref{radiuscase}.

\subsection{Measuring radius: newer methods}

\subsubsection{Differential dynamic microscopy}

DLS measures diffusion via determining the intermediate scattering function (ISF), which is the spatial Fourier transform of a time-dependent density-density correlation function\cite{Bombannes,berne76,Chu07}; it requires a laser and bespoke electronics (a correlator). Recently, a method for measuring the ISF has been demonstrated\cite{Cerbino08} that requires only the use of everyday laboratory equipment, viz., a white-light optical microscope and a CCD camera. This method, differential dynamic microscopy (DDM), exploits the fact that the intensity of a low-resolution microscope imaging is linearly related to the density of particles in the sample being imaged\cite{Wilson2011}. Thus, correlating the Fourier transform of the images gives directly the ISF.  

\subsubsection{Particle tracking}
\label{tracking}

Being a scattering method, DLS works in reciprocal space. DDM uses microscope images, but also yields the ISF in reciprocal space. In both cases, the measured quantity is the diffusion coefficient, which controls the mean-squared displacement (MSD) of Brownian particles: $\langle \Delta r^2(\tau) \rangle = nD\tau$, with $n = 2, 4$ or 6 in 1, 2 or 3 dimensions respectively. Direct real-space methods for measuring the MSD are increasingly popular. The motion of particles in a dilute sample ($\phi \ll 0.01$) can be captured by video microscopy\cite{habdas02}, and the particle motion tracked\cite{crocker96} using publicly available software\cite{idlwebsite}. Provided the microscope has been properly calibrated, such tracks yield the MSD. 

\begin{figure}
\begin{center}
\includegraphics[width=7cm]{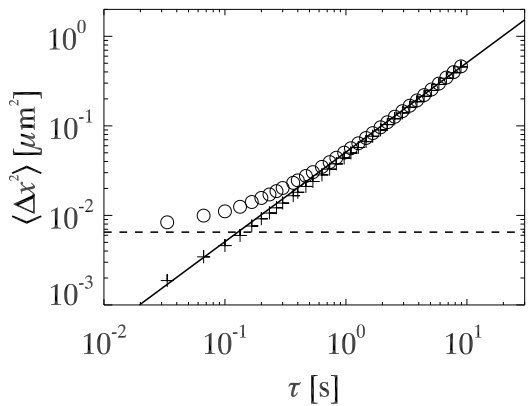}
\end{center}
\caption{
Mean square displacement of $a_C = 1.5$~$\mu$m
polystyrene spheres in a water-glycerol mixture.  ($\circ$): raw data;  (+): $\langle
\Delta x^2 \rangle - X^2$ where the estimated noise level (dashed line)
is $X=0.08$~$\mu$m.  Solid line: a linear fit to the
raw data in the range $10^0 < \tau < 10^1$~s, giving
$D=0.0254$~$\mu$m$^2$/s.  With the noise subtracted off, a linear fit to all of the data at $\tau < 10^1$~s gives $D=0.0248$~$\mu$m$^2$/s.  In this experiment, $N=9$ and $M=0.64$~$\mu$m/pixel, so the estimated
noise level $M/N = 0.07$~$\mu$m is comparable to the observed
$X=0.08$~$\mu$m.} 
\label{figmsd}
\end{figure}

Problems can occur at short and long times. The issue at short times is measurement error due to pixellation. The pixellation error for a particle whose image is $N$ pixels in diameter with individual pixels of width $M$ is roughly $M/N$. In 1 dimension, a positional
uncertainty of $X$ generates an apparent MSD of $X^2$, which is time independent, giving 
\begin{equation}
\langle r^2_{\rm meas}(\tau) \rangle = \langle r^2_{\rm true}(\tau)
\rangle + X^2.
\end{equation}
If the short-time MSD
plateau due to $X^2$ is observed, it provides an excellent means of determining
the measurement uncertainty $X$.  Figure~\ref{figmsd} shows that it is important to take this term into account for accurate determination of $D$ by tracking. Note that
changing the parameters used to identify particle positions
can often influence $X$, for better or worse \cite{crocker96}.

At long times, the measured MSD can become non-linear due to particles disappearing from the field of view, either because they leave laterally or because they become defocussed.  Since the MSD at any $\tau$ is computed based only on particles which have been observed for at least as long as $\tau$, too few particles may contribute at large $\tau$ for proper averaging.

\subsubsection{Confocal microscopy}
\label{confocal}

By using a pinhole to reject out-of-focus light, laser confocal microscopy is capable of generating images deep inside ($\sim 100 \mu$m) a concentrated suspension of fluorescent colloids. Thus, by processing images taken scanning through a sample, a 3 dimensional image of many thousands of particles can be reconstructed and their coordinates obtained\cite{idlwebsite,prasad07}. In a sample where particles are touching,  the peak of the calculated radial distribution function, $g(r)$, gives $a_{CS}$. Touching particles can be generated in a spun-down sediment\cite{kurita10rcp}, or by inducing a very short range attraction (e.g. by adding small non-adsorbing polymers). While the first peak of $g(r)$ is typically averaged over $\gtrsim 10^8$ correlations, and so is in principle highly accurate, the particle size in a typical confocal image is $\lesssim 10$ pixels, which limits the best accuracy to 
$\sim 0.1$~pixel. Furthermore, microscopy is subject to certain systematic errors \cite{royall07,baumgartl2005} that affect the determination of $g(r)$. Finally, note that the scanning mechanism drifts, so that calibration on the day of measurement is important. 

\subsubsection{Holographic microscopy}
\label{holographic}

A collimated laser
beam directed through a microscope objective scatters off a particle.  The scattered
and unscattered beams interfere in the focal plane to form a hologram.  At low enough $\phi$, the holographic image of a single, optically homogeneous particle can be fitted using Lorenz-Mie theory to determine its
position, size, and refractive index\cite{sheng06,lee07holographic,cheong09}.  This has been
demonstrated with $\sim 100$~nm to 10~$\mu$m particles.
The radius $a_C$ of an {\it individual} particle can be measured
to $\pm 10-30$~nm from a single snapshot
\cite{lee07}.  Multiple measurements further improve on this \cite{cheong09}. The sizing of core-shell particles has not yet been attempted. Note that while this method fails for exactly index-matched particles, a mismatch of as little as $1\%$ is sufficient to render it usable (D. Grier, personal communication). 

\subsection{Measuring radius: case study}
\label{radiuscase}

\begin{figure}
\begin{center}
\includegraphics[width=7cm]{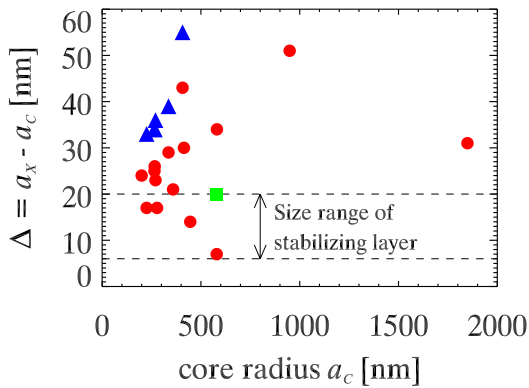}
\end{center}
\caption{
Radii of different batches of fluorescent PMMA particles determined by various methods\cite{bosma02}: $\bullet$ static light scattering, $\blacktriangle$ confocal microscopy, $\blacksquare$ $g(r)$ peak.}
\label{radii}
\end{figure}

To illustrate the difficulties in pinning down a value for {\em the} radius of a batch of particles, we reproduce data for 15 preparations of fluorescent PMMA colloids by Bosma et al.\cite{bosma02}. `Wet' particles (suspended in hexane) were characterized by SLS (far from index matching, so that $a_{CS}$ is measured), while dried particles were sized by EM. In some cases, direct measurement of sizes from confocal micrographs was performed. The peak of $g(r)$ from confocal microscopy of a close-packed sample was also reported for one sample. We take the EM radius, always the smallest, to be $a_C$. In Fig.~\ref{radii}, we plot $\Delta = a_X - a_C$, where $a_X$ is the radius from method {\em X}. The likely length of PHSA `hairs' is $6 \pm 1$~nm\cite{Cebula,Barsted}, although oligomers of up to 15-20~nm may be present\cite{Barsted}. The hairs therefore account for the lower bound of $\Delta \gtrsim 10$~nm. Larger $\Delta$ values are likely due to particles swelling in hexane, with larger particles swelling more (so that in most cases the swelling is $\sim 4\%$ of $a_{CS}$). 

The direct confocal measurements are consistently higher than the SLS data by $\gtrsim 10$~nm. This illustrates the difficulty of direct measurements from any optical image: the image of a single particle is far from sharp at the edges, both due to geometric and diffraction effects. The measurement from $g(r)$ is likely more accurate, since it relies on locating particle centres rather than edges. It is not clear why the one example of such measurements shown in Fig.~\ref{radii} is also significantly higher than the SLS result. Overall, these data illustrates that particles sizes quoted in the experimental literature may be subjected to significant systematic uncertainties that are often not always reflected in the (statistical) error bars. This must be taken into account if the particle size is then used in calculating $\phi$. 

\section{Measuring polydispersity}
\label{polysec}

Polydispersity in general refers to the existence of a distribution of particle properties, such as size, shape, charge, magnetic moment, etc. Hard sphere colloids have a distribution of radii, $P(a)$, for which we define the polydispersity, $\sigma$, as the standard deviation of this distribution divided by the mean:
\begin{equation}
\label{polyeqn}
\sigma = \left(\langle a^2 \rangle - \langle a \rangle^2\right)^{1/2} / \langle a \rangle.
\end{equation}
Very monodisperse PMMA has polydispersities
approaching 3\%, however 5-6\% is typical \cite{bosma02} for `monodisperse' PMMA. Note that some particles, including PMMA, frequently display a bimodal distribution due to secondary nucleation, so that a full distribution is needed to characterize them. 

\begin{figure}
\begin{center}
\includegraphics[width=7cm]{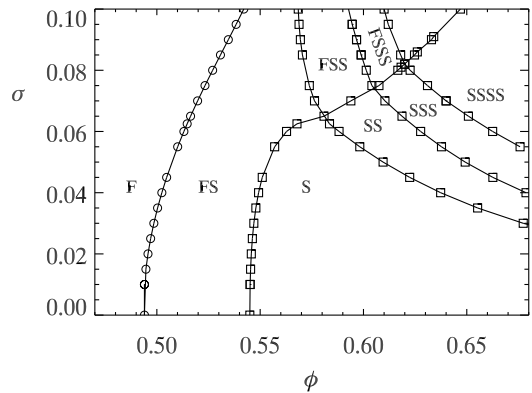}
\end{center}
\caption{
The theoretical phase diagram of hard spheres at different polydispersities, $\sigma$. F = fluid, S = (crystalline) solid; thus FSS denotes fluid-solid-solid coexistence. Replotted from\cite{Wilding2011}.}
\label{sollichfig}
\end{figure}

Polydispersity is relevant here because it affects the equilibrium phase diagram. Monodisperse hard spheres freeze at $\phi_F = 0.494$ to form crystals at the melting point $\phi_M = 0.545$. These two values are often used as fixed points for determining $\phi$ in experiments (see Section~\ref{phases}). Theory \cite{fasolo03,Sollich10a,Wilding2011} and simulations \cite{bolhuis96,Sollich10a,Wilding2011} show that even small $\sigma$ may shift $\phi_M$ and/or $\phi_F$ significantly, Fig.~\ref{sollichfig}. Indeed, particles with $\sigma$ higher than some terminal value $\sigma^*$ will fail to crystallize at all experimentally or in simulations, although theory\cite{fasolo03} predicts phase separation into coexisting solid phases. Simulations\cite{pusey09b} predict that $\sigma^* \approx 7\%$, consistent with early experiments\cite{Pusey1987}. Determining polydispersity is therefore important for measuring $\phi$. Note that the whole distribution and not just its variance may matter, e.g. in determination nucleation rates\cite{schope07}.  

All the methods reviewed in the last section can potentially yield information on the size distribution. Direct imaging, EM or optical microscopy, can estimate the full $P(a)$, subject to the same caveats already discussed. Moreover, larger particles may swell more (Fig.~\ref{radii}; see also \cite{cheong09}), giving a correlation between size and shrinkage upon drying, so that wet and dry $P(a)$ may be different.

In DLS or XPCS, the ISF from a hypothetical monodisperse suspension decays exponentially with time. Polydispersity turns the ISF into a sum of exponentials. In static scattering, monodisperse particles give sharp minima in the form factor, which are smeared out by polydispersity. (Note that multiple scattering has the same effect, and so can masquerade as polydispersity.) In principle, these features can be fitted  to yield $P(a)$\cite{Bombannes}, subject to all the usual problems and uncertainties associated with solving an inverse problem. For DLS (or XPCS), there are well known algorithms such as CONTIN \cite{Brown1993} for backing out $P(a)$ via the distribution of decay times in the ISF. Or, less ambitiously, cumulant analysis\cite{frisken01} can be used to extract $\sigma$. Form factors from static scattering are seldom inverted directly to yield $P(a)$. Instead, one assumes, say, a Gaussian form, and the scattering profile from Mie theory is fitted to obtain $\langle a \rangle$ and $\sigma$. 
The effect of small polydispersities (a few \%) on the ISF and form factor can be treated analytically \cite{Pusey1984}, and becomes independent of the form of $P(a)$ as $\sigma \rightarrow 0$. The resulting expressions can be used to fit dynamic or static scattering data to yield rather accurate values of $\sigma$. 

\section{Measuring volume fraction}
\label{volsec}

We now turn to describe and evaluate a number of methods for determining the volume fraction of model colloids. 

\subsection{Measuring mass and density}
\label{weight}

The method used by Pusey and van Megen to determine $\phi$ in their classic work on hard-sphere colloid phase behaviour\cite{Pusey87a} and described subsequently in detail in a symposium paper\cite{Pusey87b} remains conceptually the simplest. They dried a suspension of known total (or `wet') mass to determine the mass of dry particles, and converted the resulting mass fraction into $\phi$ using literature values of the densities of the solvents and of (dry) PMMA. There are multiple assumptions behind this procedure that lead to systematic uncertainties. In particular, this procedure assumes that the properties of dry and wet particles are the same, which, due to solvent absorption and solvation of the `hairs', is unlikely to be true. Thus, many have subsequently proceeded differently. If $\phi$ has been separately determined for one sample using other methods (e.g. at $\phi_F$ or $\phi_M$, see below), the the ratio of mass to volume fractions can be used to calibrate other samples. But the exact relationship between these two quantities is not a direct proportionality, and involves (unknown) ratios of the properties of wet and dry particles. 

A somewhat more involved procedure is in principle less problematic\cite{Sloutskin}. First, one determines the hydrodynamic radius $a_H$ from dynamic light scattering. Then the sedimentation velocity, $v_s$, of a dilute suspension is determined by analytic centrifugation to obtain the density difference between the (wet) particles and the solvent, $\Delta \rho = \rho_p^{\rm wet} - \rho_s$: $v_s = 2\Delta \rho g a_H^2/9\eta_0$, where $\eta_0$ is the solvent viscosity (separately measured) and $g$ is the gravitational acceleration, although the assumption that the non-slip boundary condition holds at the `hairy' particle surface may not be strictly valid (E. Sloutskin, personal communication). Since liquid densities can be determined very accurately using pycnometry or other densitometric methods if the temperature is controlled, we can measure $\rho_s$ and the density of an arbitrary suspension, $\rho$, from which its $\phi$ can be determined using
\begin{equation}
\rho = \rho_s + \phi\Delta \rho. \label{density}
\end{equation}

\subsection{Measuring phase behaviour}
\label{phases}

A popular method of calibrating $\phi$ relies on the known phase behaviour of hard spheres. In particular, in the region $\phi_F = 0.494 < \phi < \phi_M = 0.545$, hard spheres show coexistence of fluid at $\phi_F$ and crystals at $\phi_M$. The fraction of crystals, $\chi$, increases linearly from 0 to 100\% over the interval.   Measuring $\chi$ for a sample within the coexistence region then gives its $\phi$. To determine $\chi$ accurately, one needs to take into account the compression of the crystalline sediment by its own weight\cite{Paulin1990}.

The main uncertainty associated with using phase behaviour to calibrate $\phi$ is the effect of polydispersity. All calculations and simulations to date agree that finite $\sigma$ increases $\phi_F$ and $\phi_M$. Thus, e.g., in the `moment free energy' calculations shown in Fig.~\ref{sollichfig}, $\phi_F = 0.5074$ and $\phi_M = 0.5540$ at $\sigma = 5\%$, the latter being a representative value of a typical preparation of PMMA colloids. To date there has been no independent experimental check on such theoretical predictions, one of the main issues being the measurement of $\phi$ in polydisperse colloids! 

Nevertheless, these results may throw light on one of the puzzles remaining from the original work of Pusey and van Megen\cite{Pusey87a,Pusey87b}, who found that if they assumed a freezing point of $\phi_F = 0.494$, their measured melting point was $\phi_M = 0.535$. The ratio $\gamma$ of these two values, which characterises the width of the coexistence gap, is $\gamma_{\rm PvM} = 1.083$. For monodisperse colloids, $\gamma_{\sigma = 0} = 1.103$, while calculations\cite{Wilding2011} for $\sigma = 5\%$ gives $\gamma_{\sigma = 5\%} = 1.092$. It is therefore possible that the narrowing of the coexistence gap observed by Pusey and van Megen is largely due to polydispersity. Note, however, that the phase diagram likely depends on the whole $P(a)$ and not just $\sigma$. 

An additional source of uncertainty is residual charge\cite{yethiraj03,royall06} so that, hard-sphere phase behaviour no longer obtains: crystallization is expected at lower volume fractions (for PMMA, see, e.g.\cite{hernandez09}). In these cases the
phase behaviour cannot be matched to that of hard spheres at all.
However, by adding salt, the charges can be screened and hard
sphere behaviour recovered to an extent (e.g., for PMMA, see\cite{yethiraj03}). 

We mention that confocal microscopy of a sample in the fluid-crystal coexistence region can be used to deduce a value for $a_{\rm eff}$ by assuming particular values for $\phi_F$ and $\phi_M$\cite{hernandez09}, subject to all the above-mentioned caveats and uncertainties. 

\subsection{Centrifugation and sedimentation}

Perhaps the quickest way to obtain samples with approximately calibrated $\phi$ is by centrifuging to obtain a sediment that one assumes to be at `random close packing' (RCP), and therefore some known $\phi_{\rm RCP}$, which can then be redispersed with fixed volumes of solvent to give samples at lower concentrations. The method can be applied even with charged particles, since hard centrifugation can reduce even such particles to a mutually-touching amorphous state\cite{kurita10rcp}.

The main problem with this method is that the theoretical status of RCP is still debated, with different simulation algorithms giving different results\cite{torquato00,hermes10}. Experimentally, different regions of the centrifuged sediment have somewhat different concentrations ($\phi = 0.60-0.64$ in silica colloids\cite{Blaaderen1995}), and little is known about the almost-certain dependence of sediment structure on centrifugation protocol. Moreover, the spun-down sediment is inevitably compressed, and will expand with time after the cessation of centrifugation, which introduces an extra degree of uncertainty. Finally, the dependence on polydispersity is poorly known\cite{schaertl94,hermes10,farr09}. 

But centrifugation is convenient, and if the protocol is kept constant, it can be used to produce a series of samples with highly accurate normalized concentrations, viz., $\phi/\phi_{\rm sed}$, where $\phi_{\rm sed}$ is the volume fraction of the sediment. 

Under this heading, we may mention that particles with small enough gravitational P\'eclet number\cite{dullens06,kurita10rcp} (either by virtue of near density matching or by virtue of being small) and low enough polydispersity will sediment slowly under gravity to form sedimentary crystals consisting of more or less randomly-stacked hexagonal close packed (rhcp) layers of particles. If the particles are monodisperse hard spheres, then $\phi_{\rm rhcp} = \pi / \sqrt{18} \approx 0.74$ in this sediment. Again, however, the (largely unknown) effect of polydispersity as well as any changes due to charges need to be taken into account.

\subsection{Confocal microscopy and particle counting}
\label{particleCounting}

Confocal microscopy can be used to locate the position of thousands of particles in a suspension. Thus, if the particle radius, $a$, is known, then counting $N$ particle in an imaging volume $V$ will yield $\phi$ directly using Eqn.~\ref{phidef}. Occasional particle mis-identification or missing a particle all together by the software give rise to erroneous $\phi$, so that it is important to cross-check particle positions identified against raw images. In particular, particles near the edge of images are often mis-identified, so that in practice a sub-volume only is considered. Finally, uncertainties in $a$ are magnified 3-fold or more in calculating $\phi$. This latter uncertainty is compounded by the issue of which of the possible radii (Section~\ref{radiusdefine}) one should use. 

\subsection{X-ray transmission}
\label{Xraytrans}

The intensity of X rays transmitted by a sample is given by $I_T = I_0 e^{-\mu x}$,
where $I_0$ is the incident intensity, $\mu$ and $x$ are the attenuation
coefficient and thickness of the sample. In the case of a colloidal suspension, $\mu = (1-\phi)\mu_s + \phi_C\mu_p$, where $\phi_C$ is the volume fraction of particle {\em cores}, and $\mu_s$ and $\mu_p$ are the attenuation coefficients of the solvent and particles. The negligible amount of electron density represented by sterically-stabilizing `hairs' means that they hardly contribute to the beam attenuation. X ray transmission can therefore be used to determine $\phi$ directly for model colloids such as charge-stabilised polystyrene \cite{Davis1991} or silica\cite{rutgers96}, but only the core volume fraction for sterically-stabilised particles.

\subsection{Measuring $\phi$-dependent properties}

The $\phi$-dependence of a number of material properties of hard-sphere suspensions are known either from analytic theory or highly-accurate simulations. In principle, therefore, measuring these properties can be used to determine $\phi$. Here we review three: viscosity, diffusivity and structure factor.

Einstein predicted that in the limit $\phi \rightarrow 0$, the viscosity of a hard-sphere suspension is given by $\eta(\phi)/\eta_0 = 1 + (5/2) \phi$, with $\eta_0$ being the viscosity of the solvent \cite{einstein1906a}. Thus, in principle, measuring $\eta(\phi)$ is a method for determining $\phi$ (e.g.\cite{deKruif85}). While suspensions in general shear thin, this should not be a problem in the very dilute limit. But temperature control is important, since $\eta_0$ is temperature sensitive (cf. Section~\ref{established}). 

The problems associated with this method have been detailed before\cite{poon96}. In essence, very low $\phi$, certainly $\lesssim 0.02$, must be reached for the Einstein result to be valid; otherwise, second\cite{batchelor77,brady95} and higher order term in this `virial' expansion needs to be taken into account. In the case cited\cite{deKruif85}, using the Einstein relation at $\phi \approx 3\%$ leads to an error in $\phi$ of $\approx 7\%$\cite{poon96}. The difficulty, of course, is that in the limit $\phi \rightarrow 0$, {\em very} accurate viscometry is needed to distinguish the dilute suspension from pure solvent. Using the Einstein relation to calibrate $\phi$ in suspensions that are too concentrated for the relation to be valid accounts for some of the spread in literature values of $\eta(\phi_F)$, the viscosity of the most concentrated stable fluid state of hard spheres. Interestingly, determining $\phi$ using the Einstein relation is strictly independent of polydispersity: in the dilute limit, each particle contributes by an additive amount that is proportional to its volume. 

Instead of measuring $\eta(\phi)$, one could determine the single-particle diffusion coefficient as a function of $\phi$. Thus, El Masri {\em et al.}\cite{elmasri09} measured the short-time self diffusion coefficient as a function of volume fraction, $D_s^s(\phi)$. The difficulty is that there are at least two different predictions for this behaviour \cite{beenakker83,tokuyama94} which leads to a
7\% absolute uncertainty in $\phi$.

Lastly, we have already mentioned (Section~\ref{established}) that analytical expressions for the static structure factor, $S(q)$, of hard spheres are available. In particular, the closed-form expression from the Percus-Yevick (PY) approximation \cite{Hansen} fits simulation data closely, provided that the empirical Verlet-Weis correction to the volume fraction\cite{VerletWeis} is applied, i.e. the PY structure factor for volume fraction $\phi^\prime$ is used for an experimental sample at $\phi$: $\phi^\prime = \phi - \phi^2/16$. Thus, fitting measured $S(q)$ can yield a measure of $\phi$, provided that the particles can be treated as hard spheres. Again, caution about residual charges applies. Alternatively, $g(r)$ determined from confocal microscopy can be fitted to the PY form or to simulation data\cite{schmidt2008} to give $\phi$. 

\subsection{Deceptive samples}

Finally, we explain how using an accurately calibrated `stock colloid' may still lead to errors in the $\phi$ of samples.  

First, we have already mentioned a number of times the issue of swelling. If particles used for calibrating volume fraction are still in the process of swelling due to solvent absorption, then samples prepared subsequently will have a higher $\phi$ than the earlier calibration would suggest. 

Secondly, preparing samples almost invariably involves transferring suspension from one container (e.g. a bottle of stock) to another (e.g. a capillary for microscopy) using (typically) a pipette or a syringe. Apart from difficulties caused by very high viscosities\cite{cheng02} and shear thickening\cite{barnes89,brown09}, there is the problem of jamming of the particles as the suspension enters a constriction\cite{haw04}, which leads to a `self filtration' effect. Particles jammed at (say) the entrance to a pipette prevents other particles from entering, but solvent continues to flow, so that the sample inside the pipette has a lower $\phi$ than the bulk suspension that we hope to transfer. Thus, a sample loaded for confocal microscopy may be more dilute than one expects. 

Thirdly, except for very well density-matched samples at a temperature accurately remaining at the temperature at which the density matching was originally achieved, suspension inevitably sediment (or cream) with time at all except $\phi_{\rm RCP}$ or $\phi_{\rm rhcp}$. This will lead to concentration gradients. Indeed, such gradients can be deliberated exploited\cite{rutgers96,kegel04,gilchrist05,martinez05,hernandez09,kurita10rcp}, e.g. to determine equations of state. But in other cases, concentration gradients lead to unintended local deviations from the average $\phi$ at which the sample as a whole was originally prepared. 

Since one of the most important uses of hard-sphere colloids is as a model to study dynamical arrest\cite{kegel00,weeks00,PhamScience2002} and associated properties such as aging\cite{courtland03,lynch08}, any of the above three sources of unintended changes in $\phi$ will have severe consequences: all suspension properties change very rapidly with $\phi$ at and above the glass transition ($\phi \gtrsim 0.58$).

\begin{figure}
\begin{center}
\includegraphics[width=5.5cm]{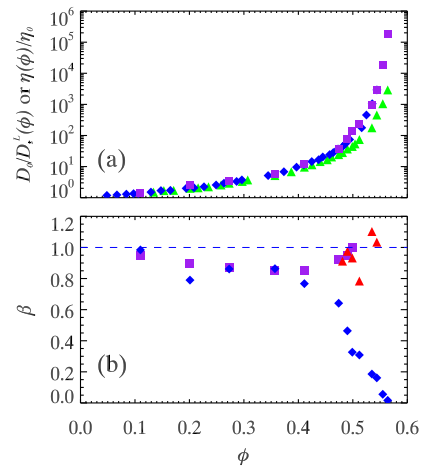} 
\end{center}
\caption{
(a) Raw data: ($\blacksquare$): The normalized long-time self diffusion coefficient of PMMA colloids as a function of volume fraction, $D_0/D_s^L(\phi)$\cite{vanmegen98}, with the volume fraction multiplied by 1.022. ($\blacktriangle$): The normalized low-shear viscosity of PMMA colloids, $\eta(\phi)/\eta_0$, as a function of volume fraction\cite{cheng02}. ($\blacklozenge$): the volume fraction of the viscosity data set being multiplied by 0.965. (b) Violation of Stokes-Einstein-Sutherland relation. The factor $\beta$, Eqn.~\ref{beta}, from the raw data ($\blacklozenge$), from theory\cite{banchio99} ($\blacksquare$) and after the $\phi$ values of the viscosity data set having been multiplied by 0.965 ($\blacktriangle$).
}
\label{caution}
\end{figure}

\subsection{Summary:  relative {\em vs.} absolute $\phi$}

The most important message from the preceding critical review is that the {\em statistical} errors involved in determining $\phi$ in the competent use of any of the above methods can almost certainly be brought below the {\em systematic} errors involved. Thus, for example, if one uses confocal microscopy to count particles and thus determine $\phi$, the major source of uncertainty is likely to be the input radius. Thus, it is perfectly possible to produce a series of samples with {\em relative} uncertainty in $\phi$ of 1 part in $10^4$. However, our collective experience in using many of these methods suggests that the systematic uncertainties are unlikely to be below 3-6\%. Far from being a small error, such uncertainties can have dramatic effects. Thus, e.g., the viscosity of a hard-sphere suspension\cite{poon96} grows by a factor of 2 when $\phi$ increases from 0.47 to 0.49; and the simulated crystal nucleation rate\cite{auer01} near $\phi_F$ can change by 10 orders of magnitude for a 1\% change in the absolute value of $\phi$.  

\section{A cautionary tale}
\label{casestudy}

In this section we give a case study to illustrate how important it is to be critical about experimental $\phi$ values by analyzing two published data sets. These data sets give as a function of $\phi$ the long-time self diffusion coefficient, $D_s^L(\phi)$\cite{vanmegen98}, and the low-shear viscosity, $\eta(\phi)$\cite{cheng02}, of PMMA colloids. At the time of publication, the diffusion data was the best and most complete available, and the viscosity data remain one of the most complete to date. The two groups came to quite different conclusions about dynamical divergence at high $\phi$. van Megen and his collaborators concluded that $D_s^L(\phi)$ diverged at $\phi = \phi_g \approx 0.58$, in a manner consistent with that predicted by mode coupling theory for an ideal glass transition. Chaikin and his collaborators, however, concluded from their $\eta(\phi)$ that there was no glass transition at $\approx 0.58$; instead, they suggested that $\eta(\phi)$ diverged at RCP, $\phi = 0.64$, according to a Volgel-Fulcher law. This controversy is ongoing (see, e.g., \cite{Brambilla2009,vanMegen2010}). We do {\em not} to enter into this discussion here; instead, we use older data to illustrate many of the issues concerned with measuring $\phi$ and using experimental data sets. These issues are, of course, pertinent for the ongoing discussion. 

The first thing to notice about these two experiments is that the reported volume fractions cannot be compared directly. Both sets of authors relied on measuring phase behaviour to calibrate $\phi$ (cf. Section~\ref{phases}), but one set of authors took into account polydispersity, and one did not. van Megen and co-workers used $\phi_F = 0.494$, the value for monodisperse hard spheres, to determine $\phi$, but cautioned that their particles had a polydispersity of $\sigma = 5\%$. The colloids used by Chaikin and his co-workers also had the same $\sigma$, and they used the simulation data of Bolhuis and Kofke
\cite{bolhuis96} to move freezing to $\phi_F = 0.505$ for this polydispersity. Interestingly, the latest analytic calculations agree closely: Wilding and Sollich\cite{Wilding2011} give $\phi_F = 0.5074$ at $\sigma = 5\%$, Fig.~\ref{sollichfig}. Thus, we multiply the $\phi$ value of the van Megen data set by a factor of $0.505/0.494 = 1.022$ to make it consistent with the Chaikin $\phi$ values. The resulting data are shown in Fig.~\ref{caution}. The measurements have been normalised, $D_s^L(\phi)$ by the single-particle diffusivity, $D_0$, and $\eta(\phi)$ by the solvent viscosity, $\eta_0$. The normalized viscosity diverges at higher $\phi$ than the normalized (inverse) diffusivity. 

At $\phi \rightarrow 0$, the solvent viscosity and the single-particle diffusivity
are related by the 
Stokes-Einstein-Sutherland relation (SESR): $D_0 = k_B T/6 \pi \eta_0 a$.
At finite $\phi$, there is no {\em a priori} reason that a generalized SESR should hold for any of the many diffusion coefficients that can be defined. So we write
\begin{equation}
D_s^L(\phi) = \frac{k_B T}{6 \pi \eta(\phi) a} \times \beta\, 
\end{equation}
where $\beta$ is a numerical factor that can be restated as
\begin{equation}
\beta = \frac{D_s^L(\phi)}{D_0} \times \frac{\eta(\phi)}{\eta_0} \label{beta}
\end{equation}
We plot in Fig.~\ref{caution}(b) (diamonds) the $\beta(\phi)$ implied by the data sets in Fig.~\ref{caution}(a). In so far as $\beta \neq 1$, the SESR is violated. 

Violation of the SESR is widely known for glass-forming systems near the glass transition. In all experimental cases known (see e.g.\cite{Sillescu1997}), $\beta >1$, i.e. the particles diffuse somewhat faster than the viscosity allows according to the SESR. The fact that $\beta$ drops very substantially below unity at $\phi \gtrsim 0.4$ in Fig.~\ref{caution}(b) is therefore surprising, and merits further analysis. 

To proceed, we turn to the work of Banchio {\em et al.}\cite{banchio99}, who have calculated various diffusivities and viscosities of hard sphere suspensions within a mode-coupling framework, and have shown that their results compared well with multiple experimental data sets. Their calculations predict that $\beta$ as defined in Eqn.~\ref{beta} hovers just below unity in the range $0 < \phi < 0.50$. Fig.~\ref{caution}(b) shows that the experimental $\beta(\phi)$ from the data plotted in Fig.~\ref{caution}(a) (diamonds) essentially agrees with theory (squares) up to $\phi = 0.35$, but start to diverge thereafter. 

Since we conclude that absolute values of $\phi$ are unlikely to be accurate to better than $3-6\%$, it is interesting to note that multiplying the volume fractions in the viscosity data set by a factor of 0.965 overlaps the two normalized data sets, Fig.~\ref{caution}(a). Not surprisingly, then, this renormalization of $\phi$ also brings very substantially better agreement in $\beta(\phi)$ in the whole range of $\phi$ covered by theory\cite{banchio99} (triangles, Fig.~\ref{caution}(b)). Assuming that there is a glass transition at $\phi_g \approx 0.58$, then this renormalization of $\phi$ also brings the direction of SESR violation in the vicinity of $\phi_g$ in line with all other known glass formers, viz., $\beta > 1$. Thus, the supposed disagreement between the two data sets is well within the range of expected uncertainties in the absolute determination of $\phi$.

\section{Conclusion}

Hard sphere colloids are now part of the accepted `tool kit' of experimental statistical mechanics. What we aim to do in this critical review is to counsel caution in comparing data from experiments against theory or simulations, because there are substantial, and probably irreducible, systematic errors in determining suspension volume fraction. This situation calls for at least three responses. First, experimentalists need to take the cue from the pioneering work of Pusey and van Megen\cite{Pusey87b} and always report exactly {\em how} they arrive at their quoted $\phi$ values, and discuss likely sources particularly of {\em systematic} errors. Secondly, experimental data sets need to be compared vigilantly against each other to reveal possible discrepancies. Finally, theorists and simulators seeking experimental confirmation of their results should not be too easily satisfied with apparent agreement, at least not until in-depth inquiry into the systematic uncertainties in $\phi$ has been carried out. 

Finally, we note that while perfect hard spheres are indeed characterized by a single thermodynamic variable $\phi$, real particles are never truly hard. Some softness in sterically-stabilized particles necessarily comes from compressible `hairs', but this becomes less significant as $a$ increase. However, it is becoming clear that for larger ($a \gtrsim 0.5\mu$m) PMMA particles, a certain degree of charging is inevitable\cite{yethiraj03,royall06,schmidt2008}, which cannot be entirely screened by salt (due to limited solubility in organic solvents). Such softness means that accurate measurement of $\phi$ alone is insufficient, and introduces further uncertainties.

\section*{Acknowledgments}

We thank P. Bartlett, J. C. Crocker, D. J. Pine, P. N. Pusey, H. Tanaka, A. van Blaaderen and D. A. Weitz for helpful discussions over many years, D.~Chen for $d\eta/dT$ data, and P. Sollich for the data in Fig.~\ref{sollichfig}. WCKP holds an EPSRC Senior Fellowship (EP/D071070/1), and performed some of the analysis at the Aspen Center for Theoretical Physics. ERW was supported by a grant from the National
Science Foundation (NSF CHE-0910707). CPR is funded by
the Royal Society.

\expandafter\ifx\csname natexlab\endcsname\relax\def\natexlab#1{#1}\fi
\expandafter\ifx\csname bibnamefont\endcsname\relax \def\bibnamefont#1{#1}\fi
\expandafter\ifx\csname bibfnamefont\endcsname\relax \def\bibfnamefont#1{#1}\fi
\expandafter\ifx\csname citenamefont\endcsname\relax \def\citenamefont#1{#1}\fi
\expandafter\ifx\csname url\endcsname\relax \def\url#1{\texttt{#1}}\fi
\expandafter\ifx\csname urlprefix\endcsname\relax\def\urlprefix{URL }\fi
\providecommand{\bibinfo}[2]{#2} \providecommand{\eprint}[2][]{\url{#2}}

\end{document}